\documentclass[a4paper,11pt]{article}
\usepackage{aaskaiid}
\usepackage{orcidlink}
\usepackage{booktabs}
\usepackage{adjustbox}
\usepackage{graphics}
\usepackage{xspace}
\usepackage{caption}
\usepackage{subcaption}
\usepackage{siunitx}
\usepackage[utf8]{inputenc}  
\DeclareUnicodeCharacter{03A9}{\ensuremath{\Omega}}
\newcommand{\hi}{H{\sc\,i}\xspace}

\newcommand{\kms}{\,km\,s$^{-1}$}
\usepackage[utf8]{inputenc}
\usepackage{newunicodechar}
\newunicodechar{☆}{\ensuremath{\star}}

\usepackage{color}
\usepackage[usenames,dvipsnames,svgnames,table]{xcolor}

\title{Studying \hi and the Cosmic Web in the Era of SKA}
\ShortTitle{H{\scriptsize I} and the Cosmic Web}

\author[1,2,3]{Hengxing Pan\orcidlink{0000-0002-9160-391X}}
\ShortName{H. Pan et al.} % shortened name list for header 
\author[4]{Martin Meyer\orcidlink{0000-0002-2838-3010}}
\author[5,3]{Madalina N. Tudorache\orcidlink{0000-0002-7288-6627}}
\author[3]{S. Lyla Jung\orcidlink{0000-0001-5512-3735}}
\author[6,7]{Maryam Arabsalmani\orcidlink{0000-0001-7680-509X}}
\author[8]{Gabriella De Lucia\orcidlink{0000-0002-6220-9104}}
\author[9]{Kristine Spekkens\orcidlink{0000-0002-0956-7949}}
\author[3]{\\Matt J. Jarvis\orcidlink{0000-0001-7039-9078}}
\affiliation[1]{National Astronomical Observatories, Chinese Academy of Sciences, Beijing 100101, China}
\emailAdd{panhengxing@nao.cas.cn}
\affiliation[2]{Guizhou Radio Astronomical Observatory, Guizhou University, Guiyang 550000, China}
\affiliation[3]{Astrophysics, University of Oxford, Denys Wilkinson Building, Keble Road, Oxford OX1 3RH, UK}
\affiliation[4]{International Centre for Radio Astronomy Research (ICRAR), University of Western Australia, 35 Stirling Highway, Crawley, WA 6009, Australia}
\affiliation[5]{Institute of Astronomy, University of Cambridge, Madingley Road, Cambridge CB3 0HA, UK}
\affiliation[6]{Excellence Cluster ORIGINS, Boltzmannstra{\ss}e 2, 85748 Garching, Germany}
\affiliation[7]{Ludwig-Maximilians-Universit\"at, Schellingstra{\ss}e 4, 80799 M\"unchen, Germany}
\affiliation[8]{INAF - Astronomical Observatory of Trieste via G.B. Tiepolo 11, 34143, Trieste, Italy}
\affiliation[9]{Department of Physics, Engineering Physics and Astronomy, Queen's University, Kingston, ON, K7l 3N6, Canada}

\abstract{Neutral atomic hydrogen plays a central role in the evolution of galaxies. Yet our understanding of how gas is accreted onto galactic disks, and the way this is governed by the cascade of processes extending up to cosmic web scales, remains poorly understood. The Square Kilometre Array has the potential to significantly advance our understanding in this field, being able to both resolve galactic disks with high column density sensitivity, while also being able to survey the large volumes needed to understand the impact of processes at the level of the cosmic web. In this chapter, we examine recent observational and theoretical progress made in this area, the potential contribution of the SKA, and needed alignment with other radio and multiwavelength facilities to advance the field.}

\begin{document}
\maketitle

\section{Introduction}

Galaxy growth requires a continuous resupply of gas through the circumgalactic medium (CGM). The CGM itself is connected to the wider intergalactic medium (IGM), tracing the complex flow of matter, energy, and metals between galaxies and their cosmic environment \citep[e.g.][]{Martin2019, Saintonge_2022}. Understanding the distribution of baryons across this galaxy–CGM–IGM structure is therefore a fundamental goal of astrophysics, critical for revealing galaxy evolution mechanisms \citep[e.g.][]{Putman_2017, Faucher2023, Pan2024}. During the process of Baryons recycle, neutral atomic hydrogen (\hi) gas serves as both the principal fuel for star formation and the most extended, cold tracer of a galaxy’s baryonic ecosystem. Yet the pathways that deliver gas to disks and how these are regulated from the CGM to cosmic-web scales remain uncertain. The Square Kilometre Array (SKA) will change this by combining high column-density sensitivity, resolved kinematics, and survey volumes large enough to capture environmental trends. 
We highlight the relevant science cases and key questions below:

\begin{itemize}

\item Impact of large-scale environment: groups to filaments

The structures observed today—such as galaxies, groups, clusters, and filaments—do not evolve independently. Instead, they are intrinsically connected, forming the remarkable "cosmic web".
In groups and clusters, stripping, tides, and starvation reshape \hi; on larger scales, filaments are predicted to channel gas and angular momentum anisotropically. A unified view must link local processing to cosmic-web supply.

Key questions:\\
How do \hi content and deficiency vary from field to filaments to dense cores and  how does proximity to \hi filaments impact gas content?\\
Can we trace gas flows along filaments and connect them to galaxy morphology? \\
How significant is the preprocessing of \hi by cosmic filaments outside the galaxy cluster virial radius?

\item Understanding gas fuelling

Gas likely reaches disks via a mix of smooth IGM accretion (including  both cold flows and shock heated gas), recycled galactic fountains, and minor mergers. Because the warm/hot CGM must pass through an atomic phase, \hi is the key diagnostic of how gas is accreted onto galaxies. 

Key questions:\\
What share of the fuelling budget comes from smooth versus satellite/recycled channels?\\
Where in galaxies, and over what timescale is this material accreted? \\
What are the signatures of accretion, and how long does it take to settle into galactic disks?\\
How do the answers to the above questions depend on environment and redshift?

\item  Assembly of angular momentum and \hi kinematic relations

Angular momentum organizes galaxy structure; the outer \hi disk records its acquisition and redistribution. Linewidths and resolved velocity fields underpin kinematic scaling relations and reveal warps, lopsidedness, and non-circular motions.

Key questions:\\
How extended are the \hi disks and what are the properties of the extraplanar \hi gas?\\
Can extended \hi kinematics constrain halo spin/concentration and reconcile observations with simulations?\\
How are the angular momentum properties of galaxies related to cosmic web environment?

\end{itemize}

\hi provides the observational leverage to connect the small-scale physics of disk fuelling to the large-scale architecture of the cosmic web, while simultaneously probing the assembly of angular momentum that governs galactic structure. By delivering deep, resolved maps across vast cosmic volumes, and by operating in concert with complementary facilities across the electromagnetic spectrum, the SKA will enable a statistically rigorous, observation-driven framework for testing physical models of how galaxies acquire, retain, and lose their gas within the cosmic web.

\section{Current status of research in the field}

\subsection{Observational} 

Since the first discovery of the cosmic web in \cite{Davis1982ApJ} and \cite{deLapparent1986}, optical observations have been the primary method for mapping the large scale structures in the Universe. Notable surveys such as the 2dF Galaxy Redshift Survey \citep{Colless_2001} and the Sloan Digital Sky Survey \citep[SDSS;][]{York2000,Doroshkevich_2004} have significantly deepened our understanding of our Universe. However, optical telescopes are only sensitive to the luminous stellar components of galaxies, thus offering a powerful but still biased view of the Universe. This limitation has been partially mitigated in radio by large-scale \hi surveys conducted with the Parkes and Arecibo telescopes \citep[e.g.][]{Zwaan_2005, jones2018}, along with the Five-hundred-meter Aperture Spherical radio Telescope  \citep[FAST;][]{Nan_2011,zhang2024}, though these surveys have limitations in resolving galaxies and reaching the distant Universe. Fortunately, the SKA L-band precursors and pathfinders (MeerKAT, ASKAP and Apertif) opened a new parameter space with the high-resolution interferometric observations \citep{Maccagni_2024}. In particular, with MeerKAT - the precursor to the SKA telescope, the studies on the relation between \hi and the cosmic web have been accelerated rapidly due to MeerKAT's unparalleled sensitivity and resolution in mapping the \hi galaxies within the cosmic web \citep[e.g.][]{Ranchod_2021, Tudorache2022, Lawrie_2025}. In addition, with its eventual 20,000 deg$^2$  coverage at sufficient angular resolution to spatially resolve most \hi disks \citep{Koribalski2020,Westmeier2022,Murugeshan2024}, the ongoing WALLABY survey on ASKAP will provide an unprecedented view of galaxy alignments with the underlying large-scale structure at low redshift before the construction of SKA is complete.

Evidence for the influence of cosmic filaments on the \hi content of \hi-detected galaxies has been elusive. \citet{Kleiner_2016} perform \hi spectral stacking to show that the \hi mass fraction of massive galaxies ($\log M_{\ast}/\rm{M_{\odot}}\geq 11$) in filaments is, on average, higher than in the field. However, no statistically significant difference in the \hi fraction has been found among the local-density-controlled low-mass galaxies ($0\leq \Sigma_{5}/\rm{Mpc^{-2}}<3$ and $\log_{10}(M_{\ast}/\rm{M_{\odot}})<11$) in their sample. Similarly, in the local Universe, \citet{lee2021properties} show that there is no clear dependence of the filament distance and the \hi mass fraction among galaxies in filaments around the Virgo Cluster (see also \citealt{yoon2025mapping}). The stellar mass of their sample galaxies is below $10^{11}\,\rm M_{\odot}$. In contrast, \citet{odekon2018effect} show that the \hi galaxies (with a stellar mass range of $8.5<\log M_{\ast}/\rm{M_{\odot}}<10.5$) in the ALFALFA survey are more \hi-deficient in filaments. 

Furthermore, \citet[][]{2022A&A...657A...9C} find the number of active galaxies at close distances to filaments to be significantly smaller than that of quenched galaxies, with an increase in \hi mass for galaxies with $\rm M_{\ast} \gtrsim 10^{9}\,M_{\odot}$ and a decrease in their  \hi deficiency with increasing distance from the filament spine. This study indicates that the observed \hi deficiency is likely due to tidal interactions or ram pressure stripping given that they find the deficient systems to be preferentially in dense regions within filaments. More recently, \citet{Luber2025chiles} find that galaxies in the COSMOS \hi Large Extragalactic Survey (CHILES) sample have higher \hi fraction in filaments ($\lesssim2\,\rm Mpc$), regardless of their stellar mass. 

The investigation on the link between angular momentum and the \hi content of galaxies in the context of the cosmic web has only been possible in the last decade due to the improvements in interferometric telescopes, specifically their larger instantaneous velocity coverage, better sensitivity and wider FOV. Studies such as \cite{BlueBird_2020} and \cite{Tudorache2022} find that on average, the spin vectors of \hi-detected galaxies are preferentially aligned with the spine of their closest cosmic filaments. However, robust observational confirmation will require larger galaxy samples.

In addition to the extensive efforts in studying the \hi properties of galaxies in cosmic filaments, there have been attempts to detect the intergalactic \hi  in the nearby filaments through \hi 21 cm emission line observations \citep[e.g.,][]{2011A&A...527A..90P, 2011A&A...528A..28P, 2011A&A...533A.122P}. \citet{2011A&A...528A..28P} reported  the identification of one \hi  source with no known optical counterpart amongst 199 detected sources in the filament joining the Local Group to the Virgo Cluster. 
In a recent study \citet[][]{2025ApJ...980L...2A} detect   numerous massive intergalactic \hi clouds in a nearby cosmic filament. The presence of these  clouds in this newly identified filament, demonstrating its gas-richness, is likely linked to its unusually geometry: exceptionally straight and narrow. Such  string-like  filaments, making up a non-negligible fraction of all cosmic filaments in the Universe, have been systematically missed in all previous studies. 
The fascinating geometry of these string-like filaments is likely due to their deep gravitational potential which, in return, should result in the accumulation of more and denser cold gas in them. This is indeed the case for the first identified string-like filament \cite[][]{2025ApJ...980L...2A}. With their remarkable straight and narrow geometries, the large-scale cold gas streams in them  are likely to have higher densities, making  them excellent candidates to empirically study the streaming of cold gas in the cosmic web. \cite{Tudorache_2025} also find a string-like filament first detected using \hi galaxies, embedded in a larger scale, optically-detected cosmic filament. Within this structure, the angular momentum of \hi gas is relatively undisturbed as its spin is strongly aligned with the optical filament.

\subsection{Theoretical}

On the theoretical front, significant progress has also been made over the past few decades in understanding the relationship between \hi and the cosmic web. Notably, the Illustris and SIMBA simulations \citep{Nelson_2015,Dav__2019} provide large cosmological volumes for probing the \hi distribution across various environments from a hydrodynamical perspective. For instance, \cite{Bah__2025} use the EAGLE and TNG100 simulations to establish how to define and characterise filaments consistently across simulations. \cite{jego2025} investigated how the star formation activity of galaxies depends on their position within the cosmic web using the SIMBA cosmological simulation from redshift $z = 3$ to $ z = 0$, and found that that the large-scale cosmic web proximity modulates star formation in star-forming galaxies through a combination of gas supply regulation and environmental processing, with different mechanisms dominating across cosmic time. Some high-resolution simulations suggest that the gas loss by ram pressure stripping is possible among dwarf galaxies in the cosmic web \citep{benitez2013dwarf,Herzog_2022}. \cite{Bulichi_2024} show that cold gas content is progressively suppressed toward the spines of filaments, with the degree of suppression increasing at lower redshifts, consistent with trends observed in sSFR in \cite{odekon2018effect}. The relative suppression of cold gas in central versus satellite galaxies, however, depends on their distance from the filament spine. It is also important to note that, in dense environments, the association of \hi with individual galaxies can be ambiguous, as \hi often traces relatively diffuse IGM or CGM rather than being largely confined to galactic disks. Semi-analytic results broadly echo these patterns. For instance, using the Galaxy Evolution and Assembly (GAEA) semi-analytic model, \cite{Zakharova2024} find that galaxies within filaments exhibit intermediate cold gas content between galaxies in clusters and in isolation. 

Critically, simulations have begun to quantify the column density regime where galaxies connect to the cosmic web. \cite{Popping2009} and \cite{Popping2015} simulated the \hi sky and demonstrated that the transition from galactic circumgalactic medium to intergalactic filaments occurs at column densities $N_{\rm{HI}} \lesssim 10^{18}$ cm$^{-2}$, below a long plateau in the distribution function between $10^{18}$ and $10^{20}$ cm$^{-2}$. Reaching this threshold is essential for directly imaging gas accretion, yet it remains largely unexplored observationally. This gas, residing in filamentary structures, constitutes the fuel for future star formation and represents a direct observational signature of the cold-mode accretion predicted to dominate gas acquisition in present-day star-forming galaxies \citep{Kere__2005,van_de_Voort_2012,Yu2024}. Interestingly, \cite{Aragon_Calvo_2019} proposes that galaxies initially accrete star-forming gas via a network of primordial filaments, directly imprinted by initial density fluctuations. When shell-crossing occurs on intergalactic scales, this pattern is disrupted. The galaxy then detaches from its filaments, ending the supply of cold gas and, consequently, its primary mode of accretion.

In the case of angular momentum, there is not a consensus on how it relates to the filaments of the cosmic web. Semi-analytic modelling indicates that low-spin halos tend to host \hi-poor galaxies, whereas high-spin halos span a wide \hi-mass range \citep{Zoldan2017}, which is relevant to the finding that the spin of dark matter halos is not random but shows alignments with nearby filaments. Simulations such as those by \cite{Aragon-Calvo_2007, Dubois_2014, Codis_2015} predict alignment between the angular momentum vector of the dark matter haloes and the filaments of the cosmic web. However, whilst \cite{Kraljic_2020} find a spin transition for a stellar mass of $\sim 10^{10} M_{\odot}$, other simulations such as \cite{Ganeshaiah-Veena_2018} find no such transition and report a preference for overall mis-alignment at all masses. The only theoretical study approaching this problem from an \hi standpoint, \cite{Kraljic_2020}, find a possible spin transition threshold in \hi mass at $M_{\mathrm{HI}} = 10^{9.5} M_{\odot}$ using the SIMBA simulation - a transition which has not been observed in observational data due to the low-number samples.

\begin{figure*}
  \centering
  \hspace{2mm}
  \begin{subfigure}[b]{0.47\textwidth}
    \includegraphics[height=0.99\columnwidth, width=0.99\columnwidth]{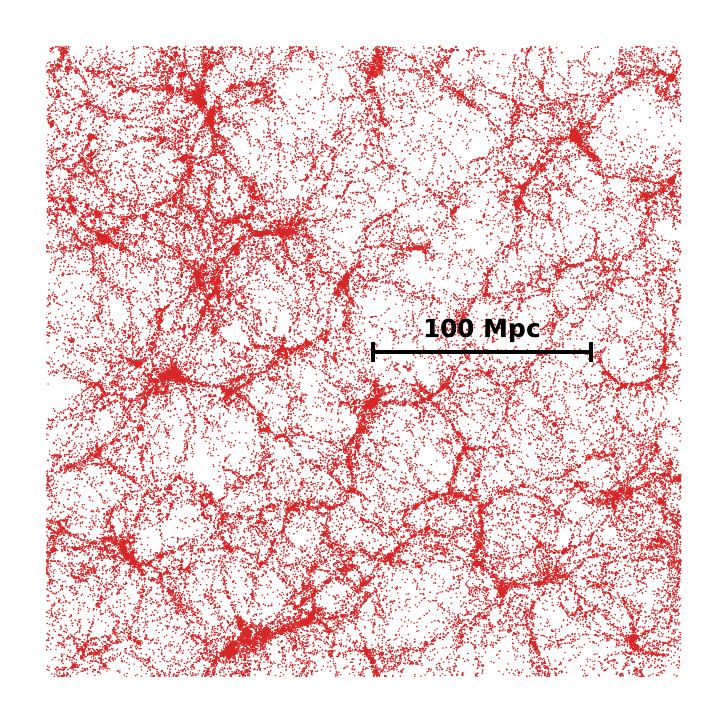}
  \end{subfigure}%
  % \hfill
  \hspace{-7mm}
  \begin{subfigure}[b]{0.47\textwidth}
    \includegraphics[height=0.99\columnwidth, width=0.99\columnwidth]{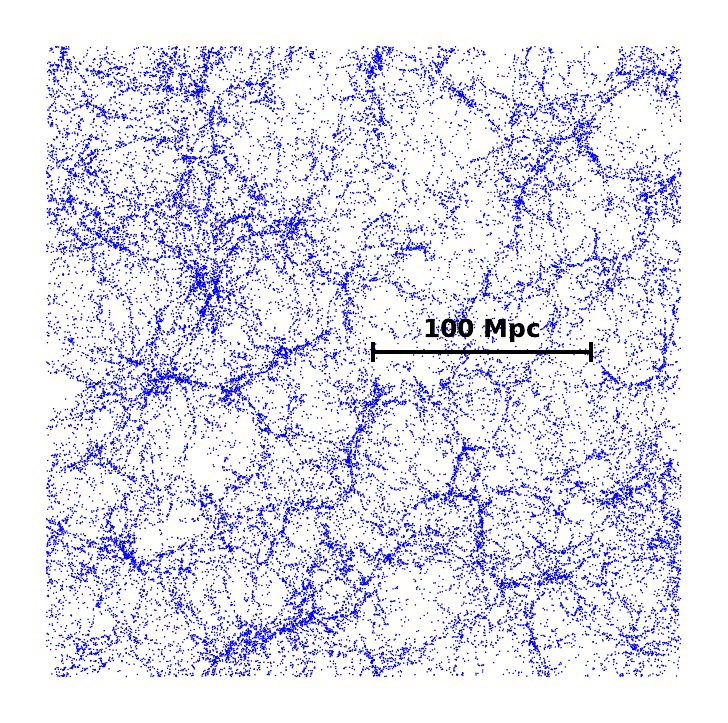}
  \end{subfigure}%
  \caption{Simulated stellar (red) and \hi distribution (blue) within a box of 300 × 300 × 30 comoving Mpc at $z=0$ in the left and right, respectively. The large-scale structures such as galaxy filaments and groups are clearly perceivable visually. The simulation is based on the IllustrisTNG-300 project \citep{nelson2021}.}
  \vspace{-5pt}
  \label{fig:cosmicweb}
\end{figure*}

We present the simulated multi-scale structures in the local Universe in Figure~\ref{fig:cosmicweb}. In the left and right panels, we show a 300 × 300 × 30 comoving Mpc simulated box of the stellar and \hi distributions at $z=0$, colour-coded by red and blue respectively, based on the IllustrisTNG-300 project \citep{nelson2021}. The large-scale structures such as filaments, groups and clusters are visually perceivable with extents ranging roughly from 100 kpc to 100 Mpc scales. The \hi galaxies appear to be less clustered than the stellar galaxies due to the diffuse nature of neutral hydrogen gas, which offers a new perspective on tracing the underlying dark matter and understanding the distribution of the cold gas derived from the dynamics of galaxy formation \citep[e.g.][]{Janowiecki_2017}.

\section{Simulated performance for SKA based on current specifications}

Based on the current SKA telescope specifications\footnote{\url{https://www.skao.int/en/science-users/118/ska-telescope-specifications}}, SKA-Mid will consist of 133 15-m SKA dishes and 64 13.5-m MeerKAT dishes at the Karoo Radio Astronomy Reserve in the Northern Cape of South Africa. The core will be composed of around 50\% of the dishes, randomly distributed within 2 km. The remaining dishes are spaced logarithmically on three spiral arms providing baselines out to 150 km. SKAO has adopted a staged approach to construction where the Low and Mid telescopes are constructed and verified in stages called array assemblies (AA), or a subset of those available (i.e. a "subarray"). We present the configuration of the SKA-Mid subarrays in Table~\ref{tab:ska}.

\begin{table}[ht]
\centering
\caption{Configuration of the SKA-Mid subarrays \citep{seethapuram_sridhar_2025_16951020}.}
\label{tab:ska}
\begin{adjustbox}{max width=\textwidth}
\begin{tabular}{lccc}
\toprule
Configuration & Number of 15-m antennas & Number of 13.5-m antennas & Maximum baseline length (km) \\
\midrule
AA4     & 133 & 64 & 159.6 \\
AA*     & 80  & 64 & 108 \\
MeerKAT & 0   & 64 & 8 \\
\bottomrule
\end{tabular}
\end{adjustbox}
\end{table}

\begin{figure}
  \centering
  \begin{subfigure}[b]{0.49\textwidth}
    \includegraphics[height=0.9\columnwidth, width=1.\columnwidth]{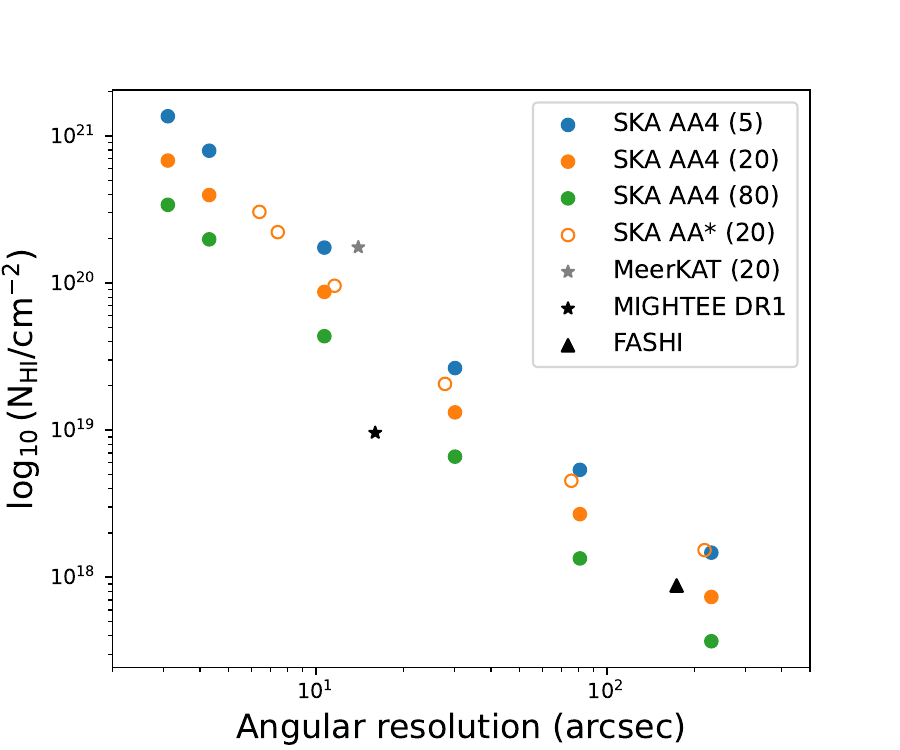}
  \end{subfigure}%
  \begin{subfigure}[b]{0.49\textwidth}
    \includegraphics[height=0.9\columnwidth, width=1.\columnwidth]{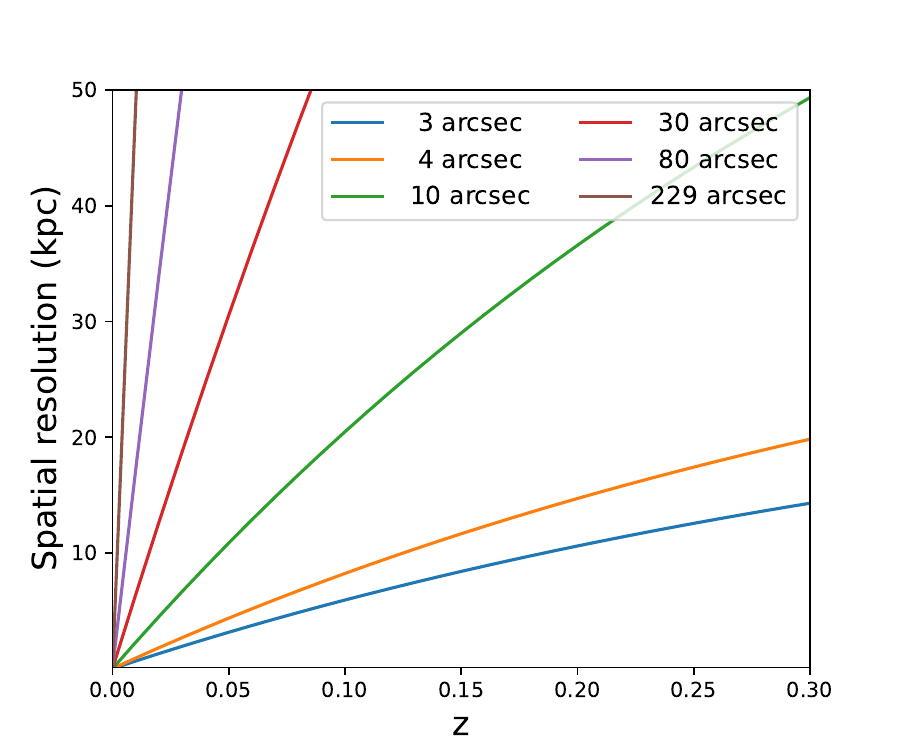}
  \end{subfigure}%
  \caption{{\it Left}: \hi column density sensitivity (3$\sigma$ over 16 km s$^{-1}$) as a function of redshift for the three mock SKA AA4 surveys color-coded by the target time of 5, 20 and 80 minutes per pointing. {\it Right}: Spatial resolution as a function of redshift for six angular resolution scales of 3, 4, 10, 30, 80 and 229 arcsec. The performance figures are estimated at a declination of -30 degrees and an observing frequency of 1.4 GHz, using Briggs weighting of robust=1 with a range of uv-tapering applied.}
  \label{fig:resolution}
\end{figure}

\subsection{Potential survey requirements and specifications}

The wide-field \hi galaxy surveys have been dominated by single-dish telescopes in the last decades until now with the Square Kilometre Array (SKA) telescope, equipped with superb sensitivity and resolution capability. The SKA will produce the largest sample of galaxies detectable in \hi emission within a reasonable observing time. Given the multi-scale nature of the cosmic web, we require large contiguous survey areas to study their morphological and statistical properties and probe \hi galaxies in various galactic environments as the Cosmological Principle states that the Universe is homogeneous and isotropic on scales above 100 Mpc based on the results of large scale surveys, hence we need to probe $\gtrsim$100 Mpc scales to fully study these multi-scale structures. On the other hand, we also need to probe $\lesssim$10 kpc scales to resolve the galaxies with great sensitivity and track their spins while they are residing in the cosmic web. To meet these requirements, we explore the \hi column density sensitivity as a function of the angular resolution for the mock SKA \hi surveys in the left panel of Figure~\ref{fig:resolution}, and plot the spatial resolution as a function of redshift in the right panel. To effectively resolve a large number of \hi galaxies at $z\lesssim0.09$, we require an angular resolution of $\lesssim$ 10 arcsec and an \hi column density sensitivity of $\lesssim10^{20}\ \rm atoms\ cm^{-2}$ (3$\sigma$ over 16 km s$^{-1}$), which the SKA AA4 and AA* can both satisfy with a target time of $\gtrsim20$ minutes per pointing.

\begin{table}
\centering
\caption{Survey specifications of SKA AA4 using the frequency Band 2 (0.95-1.76 GHz) with a channel width of 13.44 kHz (i.e. 2.8 \kms at $z=0$). The \hi column density sensitivities are estimated at 3$\sigma$ over 16 km s$^{-1}$ and the beam size is fixed at $\sim$10 arcsec. The number of sources detectable by the SKA are estimated after excluding the RFI-affected region (i.e., \(0.09 < z < 0.22\)).}
\label{tab:mock}
\begin{adjustbox}{max width=\textwidth}
\begin{tabular}{cccccc}
\hline
Survey area & Integration time& RMS & Column density sensitivity & N(z<0.09)& N(z>0.22) \\
(deg$^2$) & (minutes) & (mJy beam$^{-1}$) & (atoms cm$^{-2}$) & & \\
\hline
4000 & 5  & 0.836 & 1.74e+20 & 225002 & 51165\\                                    
1000 & 20 & 0.418 & 8.69e+19 & 86867  & 100177\\
250  & 80 & 0.209 & 4.35e+19 & 30986  & 110691\\   
\hline
\end{tabular}
\end{adjustbox}
\end{table}

Based on our experience with the SKA precursor, we assume that the maximum allocation time for a large survey project is on a scale of 1000 hours for SKA. Therefore, we fixed the total observing time at $\sim$1000 hours and simulated three surveys with [5, 20, 80] minutes per pointing corresponding to total sky coverages of [4000, 1000, 250] deg$^2$, which allows us to probe large space volumes for measuring the properties of the large-scale structures statistically without suffering substantial cosmic variance. Obviously, the deeper survey has narrower sky coverage. However, we limit our narrowest survey down to 250 deg$^2$ to maintain the necessary volume for at least partially probing the cosmic web, and in this case our mock depth is comparable to that of the MeerKAT International GHz Tiered Extragalactic Exploration \citep[MIGHTEE;][]{Jarvis2016,Heywood2024} DR1 survey. For the intermediate survey with a target time of 20 mins, we also plot the mock sensitivity of the SKA AA* configuration as a function of the resolution as a reference to the SKA AA4. At higher resolutions, the sensitivity gap between SKA AA4 and AA* narrows, indicating that AA* is already well-positioned for high-resolution observations with respect to the AA4. The sensitivity advantage of the SKA AA4 emerges at the low-resolution end, where our mock SKA surveys reach a depth comparable to that of the FAST All Sky \hi survey \citep[FASHI;][]{zhang2024} with the largest single-dish telescope of the Five-hundred-meter Aperture Spherical radio Telescope (FAST) in the world. Some key survey parameters are listed in Table~\ref{tab:mock}.

To directly observe the gas flow between galaxies and the cosmic web, a benchmark column density sensitivity of $N_{\rm{HI}} \lesssim 10^{18}$ cm$^{-2}$ is required based on simulations from \cite{Popping2009,Popping2015}. While the wide-area surveys proposed for AA4 and AA* will reach $\sim 10^{19}$ cm$^{-2}$ effectively mapping the denser circumgalactic \hi gas, they fall short of the threshold of $N_{\rm{HI}} \lesssim 10^{18}$ cm$^{-2}$ by nearly an order of magnitude. Reaching the $10^{18}$ cm$^{-2}$ regime with AA4 and AA* alone requires significant time investments on the order of 100 hours per field to resolve galactic scales in the local Universe, resulting in a limited sky coverage of $\lesssim 10$ deg$^2$ with SKA AA4 and AA*. Such studies for tracing cold gas inflow and outflow can also be found in \cite{deBlok01.2026.SKA} with more details. In this chapter, we focus on the proposed wide-area surveys to probe the multi-scale structures of the cosmic web, while also noting a powerful synergy with FAST for locating the associated gas flow. We return to this in Section \ref{sec:synergies}.

\subsection{Performance results}

\begin{figure}
  \centering
  \begin{subfigure}[b]{0.49\textwidth}
    \includegraphics[height=0.9\columnwidth, width=1.1\columnwidth]{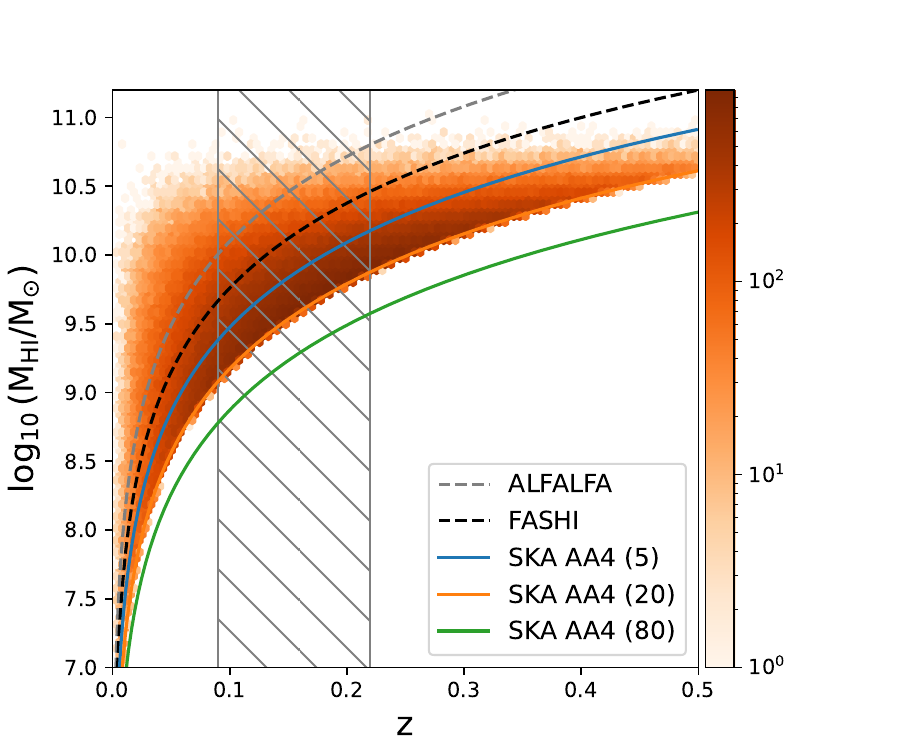}
  \end{subfigure}%
  \begin{subfigure}[b]{0.49\textwidth}
    \includegraphics[height=0.9\columnwidth, width=1.\columnwidth]{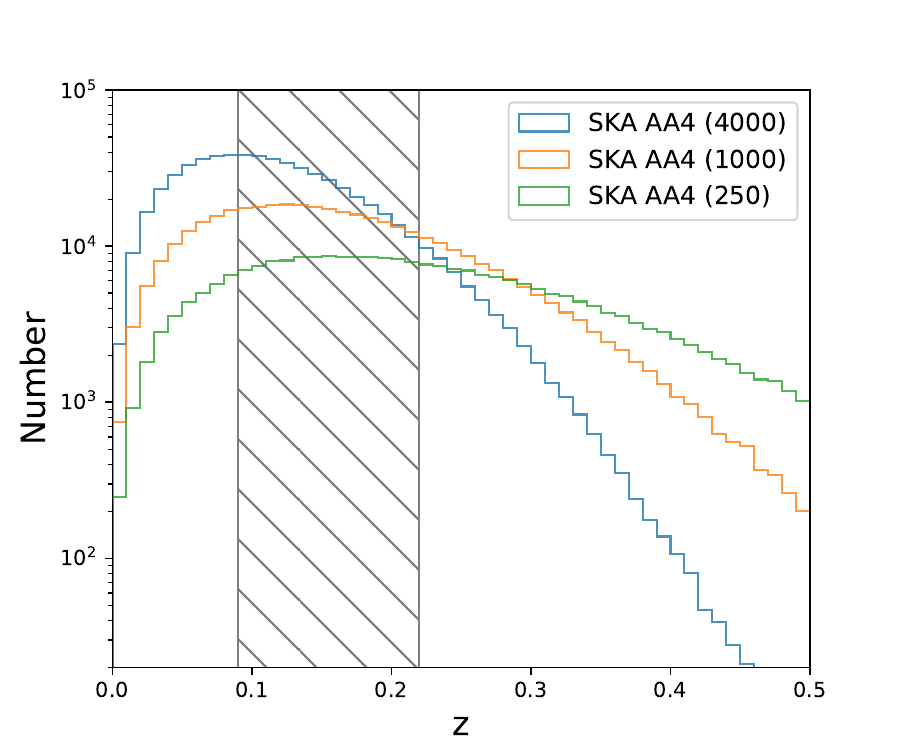}
  \end{subfigure}%
  \caption{{\it Left}: \hi mass threshold (5$\sigma$) as a function of redshift for the four mock surveys color-coded by the target time of 5, 20 and 80 minutes per pointing with blue, orange and green corresponding to an sky coverage of 4000, 1000 and 250 deg$^2$, respectively. The orange histogram shows the number of \hi detections for SKA with 20 mins target time per pointing. The grey and black dashed lines correspond to the detection thresholds of the ALFALFA and FASHI surveys \citep{jones2018,zhang2024}, respectively. {\it Right}: \hi source count as a function of redshift for the mock surveys.}
  \label{fig:threshold}
\end{figure}

We translate the SKA AA4 sensitivity at a resolution of $\sim$10 arcsec to the \hi mass threshold by assuming a typical \hi linewidth of 150 \kms, and forecast the source count based on the \hi mass function at $z\sim0$ from the ALFALFA survey. In the left panel of Figure~\ref{fig:threshold}, we show the \hi mass detection threshold as a function of redshift for the SKA AA4 mock surveys. Compared to ALFALFA and FASHI, the SKA AA4 offers a significant sensitivity advantage of roughly an order of magnitude in detecting the total \hi mass for distant galaxies for observations of 20 and 80 minutes per pointing, respectively, although we note that ALFALFA and FASHI are conducted in drift-scan mode.

\begin{figure}[ht]
    \centering
	\includegraphics[width=0.6\columnwidth]{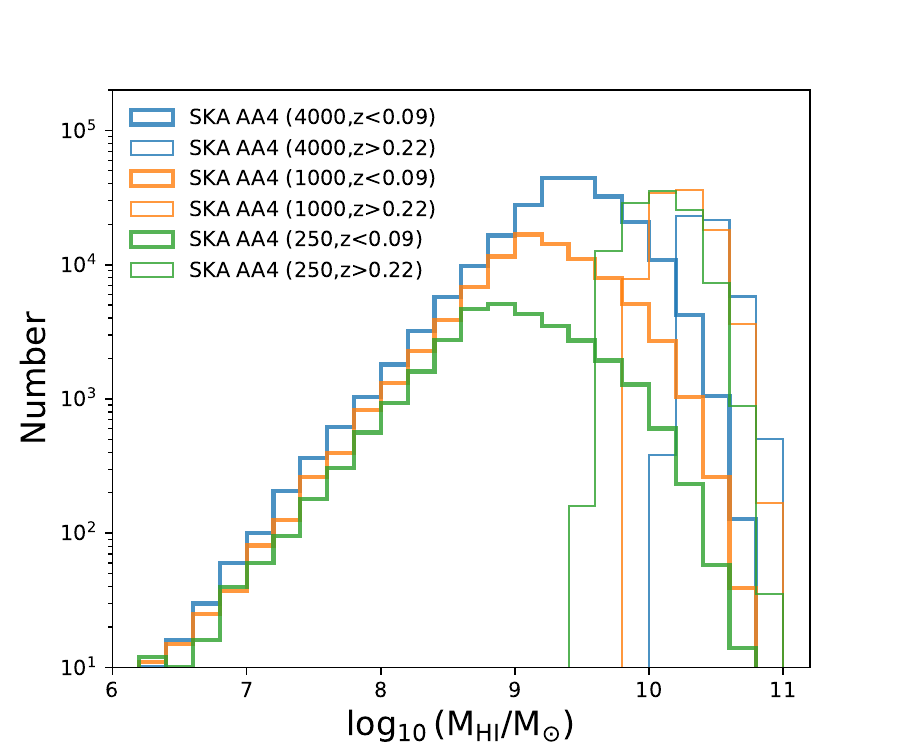}
    \caption{\hi source count as a function of \hi mass for the mock SKA AA4 surveys. The thick and thin lines are at $z<0.09$ and $z>0.22$, respectively. }
    \label{fig:mass_distribution}
\end{figure}

The source counts as a function of redshift are shown in the right panel of Figure~\ref{fig:threshold}. The differences among the three mock SKA AA4 surveys suggest that the number of \hi detections increases faster if we spread the target hours into sky coverage than into the integration depth at the low redshift range of $z<0.09$. At $0.09<z<0.22$, the RFI significantly blocks the detection rate of \hi galaxies. The total number of detections from the three mock surveys converges in the intermediate redshift range of $0.2<z<0.3$, above which the depth of the survey plays a dominant role over the sky coverage in determining the number of detections. This trend can also be perceived in Figure~\ref{fig:mass_distribution} where the wider survey can produce higher number of \hi detection across almost the full mass range at $z<0.09$ while a narrower but deeper survey has more galaxies with $9.5<\log_{10}(M_{\rm HI}/M_\odot)<10$ at the higher redshift of $z>0.22$ as expected. These mock surveys demonstrate the unique power of SKA for providing the requested sensitivity and resolution to detect $\sim$100,000 resolved (or partially resolved) \hi galaxies with 1000-hour observations. Such rich datasets allow us to address a variety of fundamental questions about galaxy formation and evolution along with the large-scale structures of the Universe.

The 4000 deg$^2$ survey covers the largest space volume that is ideal for inferring the clustering strength and bias parameters of gas-rich galaxies, and we expect a detection of Baryonic Acoustic Oscillations (BAOs) from \hi survey independently of the past and forthcoming optical surveys so to provide an \hi perspective on constraining the dark energy parameter, and the controversial Hubble constant (H$_0$). A proposal covering a much larger survey area with SKA-Mid is available in \cite{Nasirudin01.2026.SKA}. The 250 deg$^2$ survey is our deepest mock survey, enabling us to reach the most distant Universe for studying the evolution of the \hi distribution with detections up to $z\sim0.5$. The intermediate 1000 deg$^2$ survey, balancing the sky coverage and the depth of the 4000 and 250 deg$^2$ surveys, will be well placed to probe the \hi distribution from intergalactic to cosmological scales, with great potential to help us understand the co-evolution of \hi galaxies and the cosmic web.

\section{Synergies with other facilities}
\label{sec:synergies}

\subsection{Combination with single dish facilities such as FAST}

The FAST All Sky \hi survey was designed to cover the entire sky observable by the Five-hundred-meter Aperture Spherical radio Telescope (FAST), spanning approximately 22000 square degrees of declination between -14 deg and +66 deg, and in the frequency range of 1050-1450 MHz. The SKA has large common field of view with FAST surrounding the equator, which provides the ideal testbed for a thorough study of the distribution of the diffuse \hi gas in CGM and IGM by combining the sensitivity capability of FAST and the high-resolution of the SKA. By convolving the SKA-\hi data to the same angular resolution of FAST, we are able to subtract the individual galaxy contributions from the FAST intensity map, leaving the residual that is the diffuse \hi gas distributed in IGM/CGM resolved out by SKA \citep[e.g.][]{Pan2024}. We are also able to cross-correlate the FAST \hi intensity map with SKA-\hi, mitigating systemics from both instruments and give a high-SNR detection of the large-scale \hi structure due to the exceptional sensitivity of both radio instruments, and provide a complementary view of the \hi mass function from the \hi detections alone.

Furthermore, FAST's unparalleled brightness sensitivity as a single dish allows it to efficiently survey large areas to the requisite depth, identifying candidate web structures. SKA AA* and AA4 can then be deployed for targeted follow-up, providing the high angular resolution necessary to confirm the filamentary morphology, resolve the gas kinematics, and measure the accretion rate onto the central galaxy. This "discover with FAST, resolve with SKA" strategy offers the most efficient path to empirically mapping the gas flow from the cosmic web to galaxies.

\subsection{Need for multiwavelength datasets to map cosmic web and trace other baryonic components}

\hi alone cannot reveal the full baryon cycle or the geometry of the cosmic web, much of which resides in ionized, hot, or molecular phases \citep[e.g.][]{Meiksin_2009,Tejos2025}. A more complete picture requires coordinated, co-spatial datasets: optical/near-IR spectroscopy to map filaments and nodes \citep[e.g.][]{Driver2019,Malavasi2020}; UV absorption (e.g., Ly$\alpha$, O VI/Ne VIII) for cool–warm ionized gas; X-ray emission/absorption for hot halos and the Warm Hot Intergalactic Medium \citep[e.g.][]{Richter_2008,Gatuzz_2024}; (sub)millimeter CO and [C II] plus far-IR dust for the cold, star-forming phase \citep[e.g.][]{Wise2020}; thermal and kinetic Sunyaev–Zeldovich to expose diffuse ionized gas \citep[e.g.][]{Mallaby_Kay_2023}; radio continuum and Faraday rotation to trace magnetic fields; and dispersion measures from fast radio bursts to inventory otherwise "missing" baryons \citep[e.g.][]{mo2025}. When these are jointly analyzed through cross-correlations, stacking, and forward modeling with \hi emission and intensity maps, and with matched depth, area, and redshift coverage, they yield a multiphase, 3D reconstruction of how gas is funnelled along the web into halos and disks, regulates star formation, and is transformed or expelled by environment.

Even with the sensitivity of the SKA, we are precluded from detecting many dwarf galaxies beyond $z\sim0.3$. Therefore, we will employ stacking techniques to infer the \hi distribution at higher redshifts, leveraging the extensive ancillary photometric and spectroscopic datasets available (e.g., SDSS, DESI, LSST and Euclid). These ancillary data provide high-quality spectroscopic redshifts for numerous galaxies up to $z \sim 0.5$, along with multiband photometry necessary for estimating stellar properties. Stacking techniques use the known positions of galaxies from surveys at other wavelengths and allow the measurement of quantities, such as the average \hi mass \citep[e.g.][]{Rhee_2017,Chen_2021,Bianchetti_2025,Luber2025}. Further, \cite{Pan2020,Pan2021,Pan2023,Pan2025} developed a new "Bayesian stacking" technique to measure the HIMF and \hi scaling relations for far lower-\hi systems and the higher-z Universe than possible from direct detections. 

\section{Conclusions}

Advancing our understanding of gas fueling and galaxy evolution requires a precise mapping of the cosmic web and its influence on key properties such as angular momentum and \hi content. While current observational studies and theoretical simulations have begun to establish crucial correlations between galaxies and their large-scale environments, from groups to filaments, significant limitations remain in both mapping the cosmic web and modelling its physical impact. The advent of the SKA, with its revolutionary \hi survey capabilities as outlined in our performance projections, promises to overcome these hurdles. By achieving the necessary survey depth and resolution, the SKA will enable a statistically robust investigation of gas in diverse environments. Crucially, the full scientific return of the SKA will be unlocked through synergies with powerful single-dish facilities like FAST for zero-spacing data and multiwavelength campaigns that trace the full baryonic content of the cosmic web. This coordinated effort will ultimately provide a unified, multi-scale picture of how the cosmic environment governs the accretion, loss, and assembly of gas in galaxies.

\section*{Acknowledgments}

We acknowledge the use of the ilifu cloud computing facility - \url{www.ilifu.ac.za}, a partnership between the University of Cape Town, the University of the Western Cape, the University of Stellenbosch, Sol Plaatje University, the Cape Peninsula University of Technology and the South African Radio Astronomy Observatory. The ilifu facility is supported by contributions from IDIA and the Computational Biology division at UCT and the Data Intensive Research Initiative of South Africa (DIRISA). HP acknowledges support from National Key R\&D Program of China (No.~2022YFA1602904, No.~2024YFA1611502, No.~2025YFE0202300), National SKA Program of China No.~2025SKA0150100, Guizhou Provincial Science and Technology Projects (No.~QKHFQ[2023]003, No.~QKHFQ[2024]001, No.~QKHPTRC-ZDSYS[2023]003) and Guizhou Province High-level Talent Program (No.~QKHPTRC-GCC[2022]003-1). SLJ acknowledges the support of a UKRI Frontiers Research Grant [EP/X026639/1], which was selected by the European Research Council, and the STFC consolidated grants [ST/S000488/1] and [ST/W000903/1].

\bibliographystyle{abbrvnat-maxbibnames4}
\bibliography{reference} % if your bibtex file is called example.bib

\end{document}